# On-chip Single-crystal Plasmonic Optoelectronics for Efficient Hot Carrier Collection and Photovoltage Detection

*Yunxuan Zhu[1,2,*], Sai. C. Yelishala[1], Shusen Liao[2], Jackson Shropshire[3], Douglas Natelson[2,4,5,*] and Longji Cui[1,6,*]*

[1]Paul M. Rady Department of Mechanical Engineering, University of Colorado Boulder, Boulder, CO 80309, United States.

[2]Department of Physics and Astronomy, Rice University, Houston, TX 77005, United States.

[3]Department of Chemical and Biological Engineering, University of Colorado Boulder, Boulder, CO 80309, United States.

[4]Department of Electrical and Computer Engineering, Rice University, Houston, TX 77005, United States.

[5]Department of Materials Science and Nanoengineering, Rice University, Houston, TX 77005, United States.

[6]Materials Science and Engineering Program and Center for Experiments on Quantum Materials, University of Colorado Boulder, Boulder, CO 80309, United States.

*Corresponding authors: yunxuan.zhu@colorado.edu (Y.Z), natelson@rice.edu (D.N), and longji.cui@colorado.edu (L.C)

**Abstract**

Large-area chemically synthesized single-crystal metals with nanometer-scale thickness have emerged as promising materials for on-chip nanophotonic applications, owing to their superior plasmonic properties compared to nanofabricated polycrystalline counterparts. While much recent attention has focused on their optical properties, the combined optimal electrical and optical characteristics, which hold great potential for high-performance optoelectronic functionalities, remain largely unexplored. Here, we present a single-crystal plasmonic optoelectronic platform based on nanowires fabricated from synthesized gold flakes and demonstrate its capabilities for highly enhanced hot carrier collection, electroluminescence, and photovoltage detection. Notably, single-crystal gold nanogap devices exhibit an order of magnitude higher open-circuit photovoltage compared to polycrystalline devices, representing one of the highest reported photovoltage sensing performances in terms of on-chip device density and responsivity per area. Our analysis revealed that this enhancement is attributed mostly to the suppression of electron-phonon scattering and improved hot carrier tunneling efficiency in single-crystal devices. These results highlight the potential of large-scale single-crystal nanostructures for both fundamental studies of nanoscale hot carrier transport and scalable electrically driven nanophotonic applications.

**Introduction**

Metallic nanostructures play a central role in plasmonics research due to their ability to allow strong light-matter interactions and generate highly nonequilibrium hot carriers [1,2], which hold promise for a wide spectrum of applications including chemical sensing[3–6], imaging[7–9], plasmon-enhanced catalysis[10], and on-chip nanophotonics[11–14]. In recent years, synthesized large-area (>$10^3$ μm$^2$) single crystal (SC) metal flakes with thickness down to a few nanometers have attracted much interest due to their excellent plasmonic properties and the need for uniform building blocks in various nanophotonic applications[15]. In particular, unlike polycrystalline (PC) nanostructures which contain structural defects and multicrystalline domains that degrade plasmonic properties, SC nanostructures offer strong plasmonic resonances near theoretical limits [16,17], ultralow propagation loss[15,18], and enhanced optical nano focusing[19–21]. The highly crystalline, atomically smooth SC structures also improve fabrication precision, addressing the challenges of PC nanostructures such as non-uniform resistance to focused ion beam (FIB) milling[15,22], and the associated inconsistent functionality. Furthermore, compared to small-scale SC materials such as nanoparticles and nanorods[22–25], the ability to synthesize SC structures at large scales presents unique opportunities for on-chip scalable integration, especially when complex plasmonic structures with high device uniformity and low fabrication tolerances are required.

Recent progress has synthesized SC metal flakes with area up to $10^4$ μm$^2$ and thickness down to 2 nanometers [26–30]. SC Au flakes have been shown to achieve high electron injection [31] and harmonic generation efficiencies orders of magnitude higher than PC nanomaterials [30]. Furthermore, recent experiments have demonstrated various applications of SC nanostructures including nonlinear photoluminescence[32–34], bright single-photon emission[35], and photocatalysis [36,37]. Despite providing great insights, these pioneering studies have mostly focused on the optical properties, while the optoelectronic properties of SC nanostructures remain largely unexplored [38]. This missed the opportunity to leverage the combined favorable electrical and optical properties of SC nanomaterials that are both required in light emission and photovoltaic sensing applications. In particular, recent years have seen increased interests in photodetection methods operating in open-circuit photovoltage mode due to their simplicity, low power consumption, and ability to directly generate sensing signals without external loads. However, the performance of these devices is largely limited by the material compositions and device geometry [39–41]. Non-plasmonic photodetectors, such as those based on semiconductors and metals, rely on intrinsic electronic processes such as band-to-band transitions[42], photo-thermoelectric effects[43], or photoemission[44]. In contrast, plasmon-enhanced photodetectors leverage surface plasmon resonances (SPRs) in metallic nanostructures to confine light into subwavelength volumes, enhancing the responsivity and spatial resolution [45]. It is of great technological interest to discover novel physical mechanisms that can potentially further shrink the device footprint of the photodetectors while maintaining the high responsivity. In this context, simple, single-metal material-based SC nanostructures offer a pathway to potentially outperform existing photovoltage detection methods.

In this work, we present a SC plasmonic platform based on synthesized large Au flakes and demonstrate its optoelectronic functionalities including highly enhanced hot carrier collection, electroluminescence, and photovoltage detection. While our previous studies focused on the fundamental mechanisms of hot carrier dynamics in PC plasmonic tunnel junctions[46–49], this work presents a platform-level approach to hot carrier optoelectronics by leveraging SC Au for reproducible and scalable on-chip photodetectors with order-of-magnitude enhancement in photovoltage generation compared to their PC counterparts. Additionally, we benchmark device performance against a range of on-chip photodetectors, highlighting both enhanced responsivity and integration density. A practical fabrication strategy is also provided to achieve high-yield, consistent device production. Specifically, we fabricated SC Au flakes into nanowires devices, which were subsequently electromigrated to create sub-nanometer-sized tunneling gaps to significantly enhance the localized plasmon resonances and allow hot electron transport and energy collection via electrical means. In continuous metal nanowires, the photovoltage response is dominated by photo-thermoelectric effects [50,51]. However, in the SC nanogap devices, we observed a large, orders of magnitude higher increase in open-circuit photovoltage. Our analysis shows that the large enhancement is mostly due to the tunneling of energetic hot carriers generated through non-radiative plasmon damping, which is different from previous photodetectors and thus enables exceptionally high responsivity with miniaturized device footprint. Furthermore, the compact on-chip devices presented here offer a simple measurement platform to infer fundamental transport properties of plasmonic hot carriers at the nanoscale that are difficult to access previously.

**Results**

We synthesized hexagonal or triangle-shaped SC gold flakes on a $SiO_2$/Si substrate (Fig. 1) following a chemical process developed in [15] (see Materials and Methods). The thickness of these flakes ranged from ~10 to 60 nm, with the 20-nm flakes selected in our experiments due to their optimal plasmonic properties corresponding to the excitation wavelength (785 nm). As shown in the SEM and AFM scanning images shown in Fig. 1, the SC flakes exhibit sharp edges and atomically smooth surfaces, in contrast to the e-beam evaporated PC electrodes. The single crystallinity is further confirmed through electron beam diffraction (Fig. 1b), which is compared with that from the evaporated PC Au thin films of the same thickness (Fig. 1c).

To create plasmonic devices from the SC Au flakes, we applied FIB milling to pattern a nanowire array, one wire of which is shown below (Fig. 1e). The dimensions, 120-nm wide and 600-nm long, were optimized prior to our experiments through finite element simulations. Subsequently, sub-nanometer-sized tunneling gaps were created within the nanowires via electromigration break junction technique (see Material and Methods for details). Noted that this is a key step that allows us to probe a series of electrically driven plasmonic functions of SC metal nanostructures including light emission, hot carrier transport, and photovoltage sensing. During our experiments, we used controlled multistep electromigration process [47] to ensure that the zero-bias electrical conductance of the formed nanogap was between $0.05G_0$ to $1G_0$ ($G_0=2e^2/h=1/(12.9\ k\Omega)$). A representative nanogap device on the crystalline nanowire is shown in Fig. 1g, which is then used to measure the hot carrier tunneling induced photovoltage and electroluminescence. Note that the gap size shown here differs from the actual gap during measurements, as the SEM image was taken after transferring the device from 10 K in the cryostat to room temperature ambient conditions. The substantial change in temperature and environment can significantly alter both the gap size and geometry. To measure the optoelectronic response of the SC nanogap devices, we used a home-built combined electrical transport and optical spectroscopy setup to measure the light emission spectrum[14,49] and photovoltage response[52,53] (see the setup schematics in Fig. 1f). The excitation laser beam is modulated by a mechanical chopper at 307 Hz and raster scanned across the nanogap to excite the highly localized plasmonic resonance, which then non-radiatively decays to generate hot electrons that tunnel through the nanogap, resulting in a photovoltage signal. A net tunneling current through the nanogap requires broken symmetry in the generation and transport of the hot carriers. In the open-circuit configuration, the hot carrier current generates an open-circuit photovoltage (OCPV). The OCPV as a function of the laser position is measured through a voltage amplifier in combination with a lock-in amplifier synced to the chopper frequency.

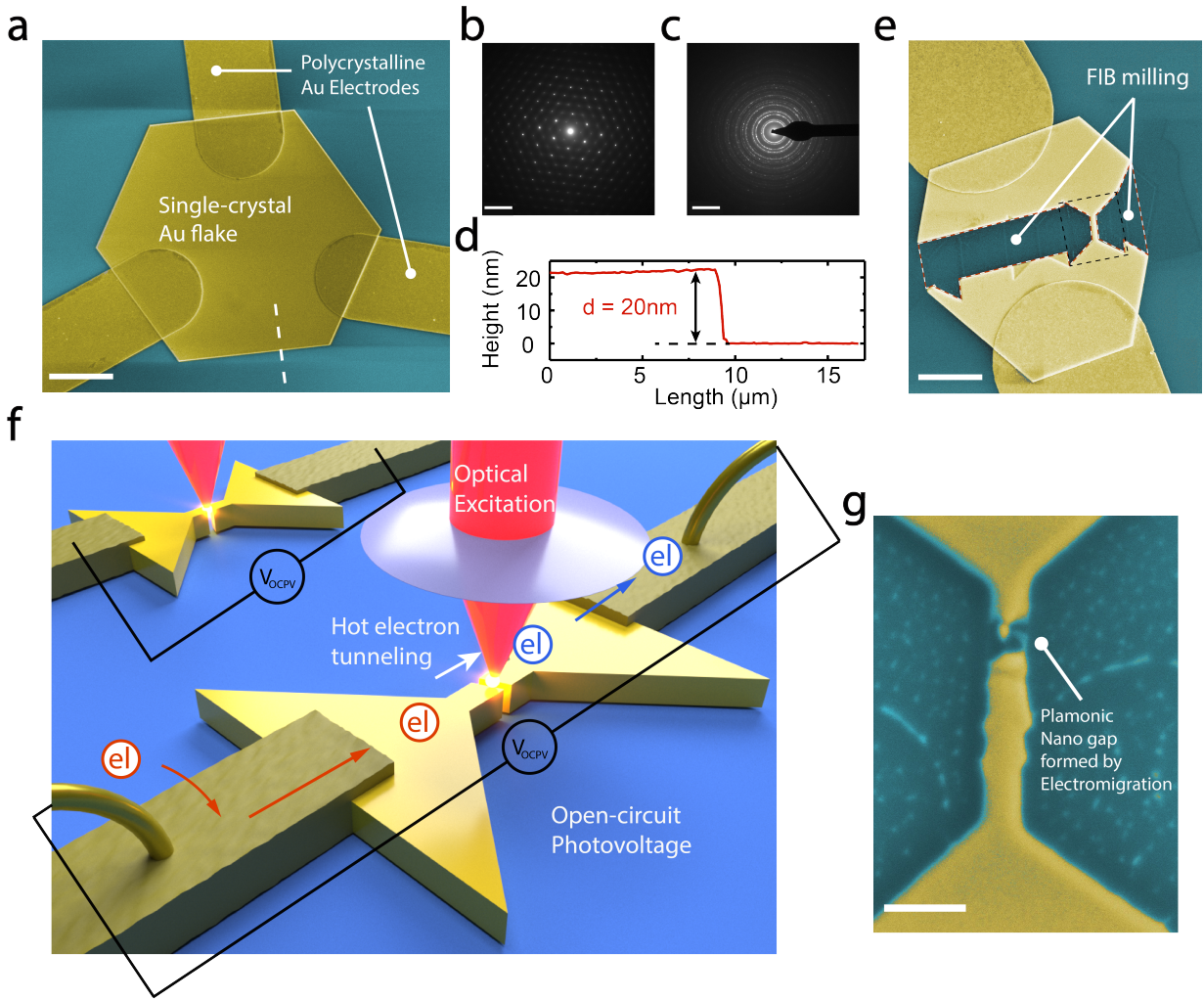

**Figure 1. Characterization for synthesized SC Au flakes and schematics of experimental setup.** (**a**) SEM images for a hexagonal SC gold flake together with deposited electrodes at the corner The scale bar here is 3.5μm. (**b**) and (**c**) Electron diffraction pattern of the 20 nm SC gold flake and 20 nm evaporated PC Au flake, respectively. The scale bar is 10 nm$^{-1}$. (**d**) AFM scanned height profile along the dased line of the 20nm thick SC flake in (**a**). (**e**) Focused ion beam (FIB) milling was used to cut a nanowire from the SC flake. The scale bar is 2.5μm. (**f**) 3D sketch of the setup. (**g**) SEM image of an electromigrated nanowire, showing a nanogap for highly enhanced local surface plasmonic responses. The scale bar is 300 nm.

As shown in Fig. 2a-c, we measured the OCPV generated in different devices made of SC Au, PC Au, and PC Au with a thin Cr adhesion layer. Cr is used due to its known properties to strongly damp the plasmonic enhancement. All three nanogap devices have a similar zero-bias resistance of ~400 kΩ. Remarkably, we observed that single crystal Au devices exhibit an order of magnitude greater OCPV than PC Au, and nearly two orders of magnitude higher than PC Au/Cr nanogaps (maximum OCPV signal is labeled on top of each map). A more detailed comparison between SC Au and PC Au is provided in the linecut data taken along the maximum OCPV signal in Fig. 2d.

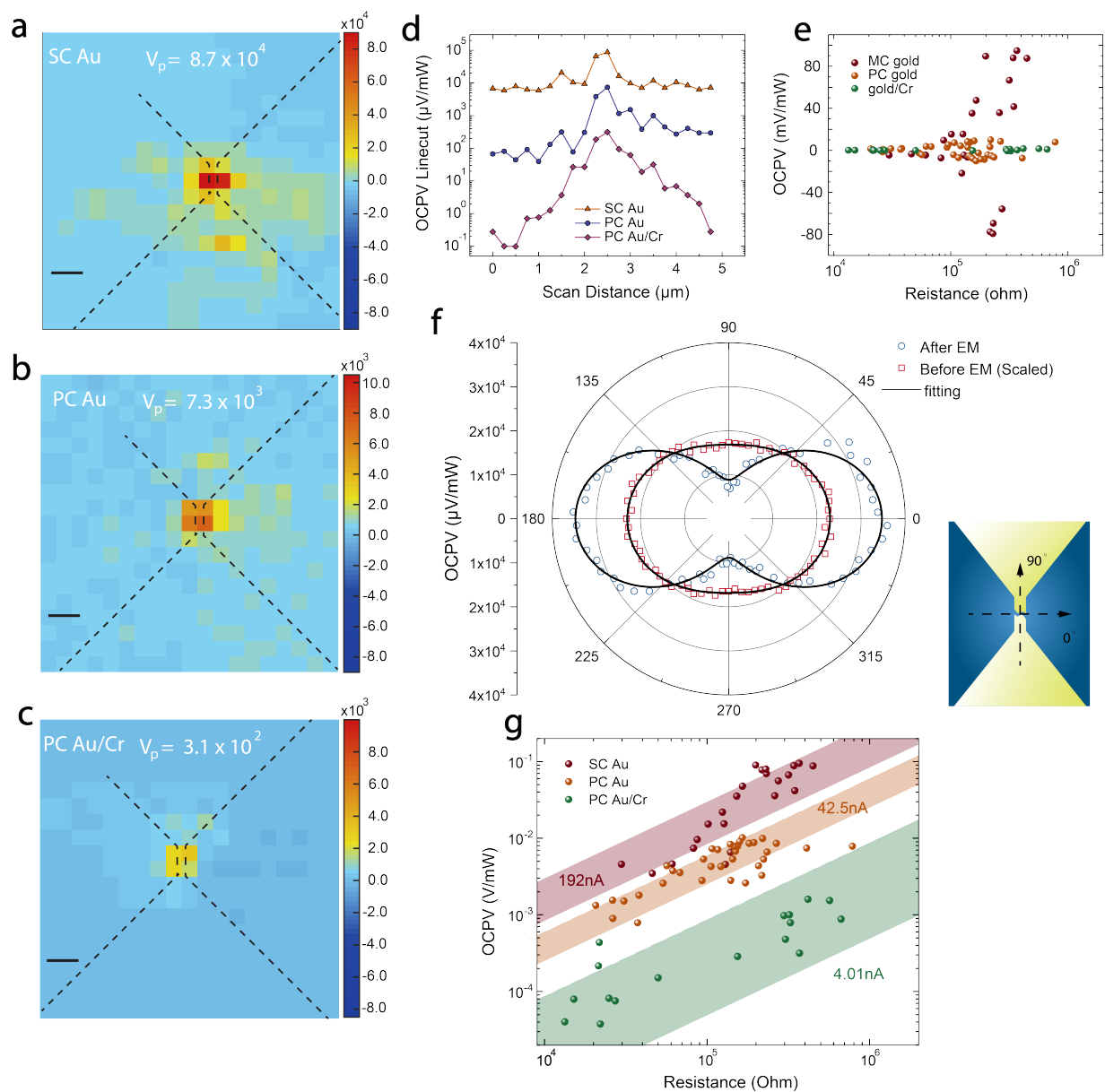

**Figure 2. Measured open-circuit photovoltage (OCPV) and analysis for three different materials. (a-c)** Scanned OCPV maps for nanogap devices made from SC Au, lithographically patterned PC Au, and PC Au with a 1 nm Cr adhesion layer as plasmon damping medium, respectively. The nanowire and bowtie shape fanout is outlined, with the maximum OCPV value indicated in the legend. The unit is µV/mW. The black scale bar in **(a-c)** is 500nm. **(d)** OCPV linecuts on logarithmic scale for three materials taken along the maximum OCPV signal in the OCPV maps shown in **(a-c)**. **(e)** Scatter plot showing the sign and magnitude of the OCPV across all the devices for different materials. Here, the positive sign indicates the hot electron tunnels towards the ground. **(f)** Polarization dependent OCPV signal before and after the formation of the electromigrated nanogap. **(g)** Statistical analysis of OCPV signal showing scatter plots on a logarithmic scale with varying zero bias conductance of three materials. The confidence intervals of hot electron tunneling currents are indicated by the shaded areas for different materials. The average values are labeled in the legends.

The polarization-dependent behavior of the measured OCPV signal reflects the plasmonic mode structure of the optically excited local and propagating surface plasmons in the nanowires with or without the formation of the nanogap (Fig. 2f). Specifically, the OCPV signal was measured as a function of the polarization of the excitation laser, after fixing the laser position at the maximum OCPV location within the map. As shown in Fig. 2f, the polarization-dependent response of the nanowire structure with a formed tunneling gap in OCPV before and after the electromigration is fitted to a cosine-squared function of the form $A\cos^2\theta + B$. The OCPV before nanogap formation is artificially amplified by a factor of 3000 for better comparison with the OCPV after the nanogap formation. Prior to the nanogap formation, the OCPV signal is very weak and exhibits minimal polarization dependence, suggesting that the response is dominated by the simple photo-thermoelectric responses of the continuous nanowire[52]. However, once the nanogap is formed, the OCPV signal becomes significantly stronger and displays a peak in OCPV when the laser polarization is orthogonal to the nanowire. This polarization dependence is consistent with previous observations, where the transverse plasmon resonance of the nanogap due to hybridization with nanogap multipolar modes[54] is optimally excited when the incident electric field is perpendicular to the nanowire. In contrast to the case of a continuous nanowire, the main source of the observed giant photovoltage is the tunneling of plasmon-induced hot carriers through the nanogap[52] (see below for detailed analysis).

To further evaluate the magnitude of hot electron tunneling, we performed a statistical analysis to quantify the tunneling current (calculated by dividing the OCPV with junction's zero bias resistance) in nanogaps formed with SC Au, PC Au, and PC Au/Cr. As can be seen in Fig. 2g, the OCPV scatter plots with varying zero bias conductance provide a detailed comparison of different materials. Each material exhibit a distinct range of hot electron tunneling currents, with confidence intervals indicated by the shaded areas. Here, the confidence intervals (95%) for each material are obtained by calculating the average and deviation in the tunneling current across different devices. For SC Au, the average hot electron tunneling current was calculated to be approximately 192 nA, showing nearly a five-fold enhancement compared to the 42.5 nA observed for PC Au, and a ten-fold increase compared to the 4.01 nA measured for PC Au/Cr. The consistent distribution of the scatter plots across multiple measurements for each material suggests that the fabrication techniques employed here yield reproducible results. We note that reproducibility of nanogap devices is crucial for developing high-performance plasmonic platform based on SC Au flakes to leverage the enhanced plasmonic optoelectronic properties of these devices. Furthermore, the sign of the OCPV across different devices, indicated by the direction of hot electron tunneling where the positive sign corresponds to tunneling towards ground, have been summarized in Fig. 2e. It can be clearly seen that a roughly symmetric distribution in the tunneling direction can be observed, consistent with the origin of the OCPV due to the hot carrier generation asymmetry in the junction[53].

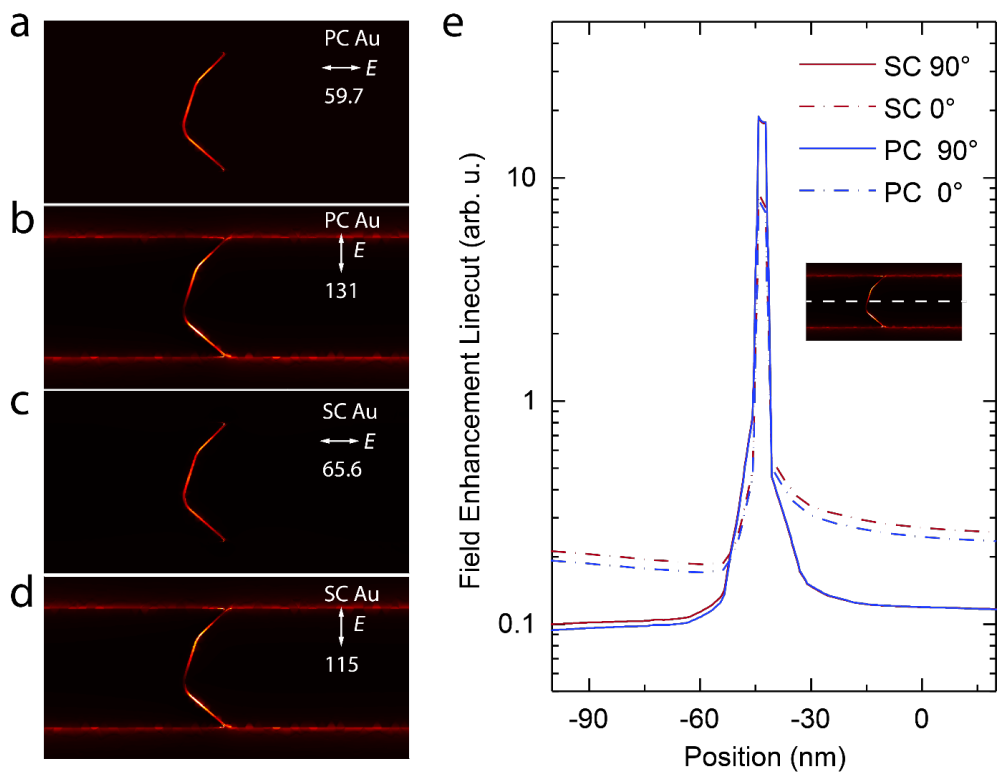

**Figure 3. Simulated field enhancement of the tunneling nanogap devices at different polarizations.** (a-d) Contour plot of the calculated plasmonic field enhancement at 785 nm photon energy for PC and SC Au nanogaps with 0° and 90° polarization of the electrical field. The field enhancement ratio has been labeled

in the plots. (e) Horizontal linecuts of the contour plots in (a-d), with the cut crossing the center of the nanogap.

To elucidate the physical mechanisms for the large enhancement of the observed open-circuit photovoltage signal, we first evaluated the plasmonic field enhancement within the nanogap and the surrounding region, which directly reflects the nanostructures' capability to generate hot carriers under optical or electrical excitation [47,49]. The calculation was performed on SC and PC Au at both 0- and 90-degree polarizations (see Supporting information for details). As shown in Fig. 3a-d, the results show that the plasmonic field enhancement for SC Au nanogaps is comparable to that of PC Au nanogaps, with the 0 degree polarization data exhibiting roughly twice the field enhancement of the 90 degree polarization, consistent with the observed OCPV polarization dependence shown in Fig. 2d. Note that the dielectric functions of SC and PC gold used in the simulation were taken from a reference where the SC gold quality does not match that of the chemically synthesized flakes used in this work[55]. Additionally, the precise geometry of the tunneling gap can also influence the field enhancement. While we made slight adjustments to both the gap structure and the dielectric parameters, the overall conclusion that SC and PC gold exhibit comparable field enhancements remains robust.

The results in Fig. 3 suggest that both SC Au and PC Au nanogaps have similar plasmonic field enhancement performance, which cannot explain the dramatic difference in the hot electron tunneling current in these two nanomaterials. In light of this finding, we proceed to investigate the transport properties of the optically excited hot carriers. The excited surface plasmons undergo rapid nonradiative damping and eventually generate hot electrons and hot holes far away from the Fermi level. To provide further insights we performed electroluminescence (EL) measurements of the plasmonic nanogap devices by applying an increasing bias voltage across the nanogaps while measuring their light emission spectra. As shown in Fig. 4a, the photon emission intensity of EL spectra for a typical SC Au nanogap device increases with higher bias. It has been shown in previous work that these spectra provide valuable information on the steady-state distribution of hot carriers generated by the non-radiative decay of electrically excited plasmons when a high electrical current is supplied to sustain the transport of hot carriers [14,48,49]. A normalization method developed in the previous work [14,48,49] was used to extract the effective electron temperature ($T_{eff}$), which characterizes the energy of the Boltzmann-like steady state distribution ($e^{-\hbar\omega/k_B T_{eff}}$) of plasmon-induced hot carriers [46,49]. This method involved dividing the spectra taken at different bias by a reference spectrum acquired at a fixed bias, to separate the contribution from bias-dependent hot electron dynamics[48]. The normalized spectra are shown in Fig. 4b, where all the reduced spectra show a linear decay behavior with photon energy on a logarithmic scale, enabling the effective temperature to be inferred through Boltzmann distribution fitting. Furthermore, by incorporating the Boltzmann factor into the measured spectra, the bias-independent plasmonic local density of states (LDOS) for this nanogap device can be determined. As shown in the inset of Fig. 4a, it is noted that all normalized spectra collapse into a single curve, indicating that the nanogap geometry of the SC Au tunnel junction has remained the same under different biases.

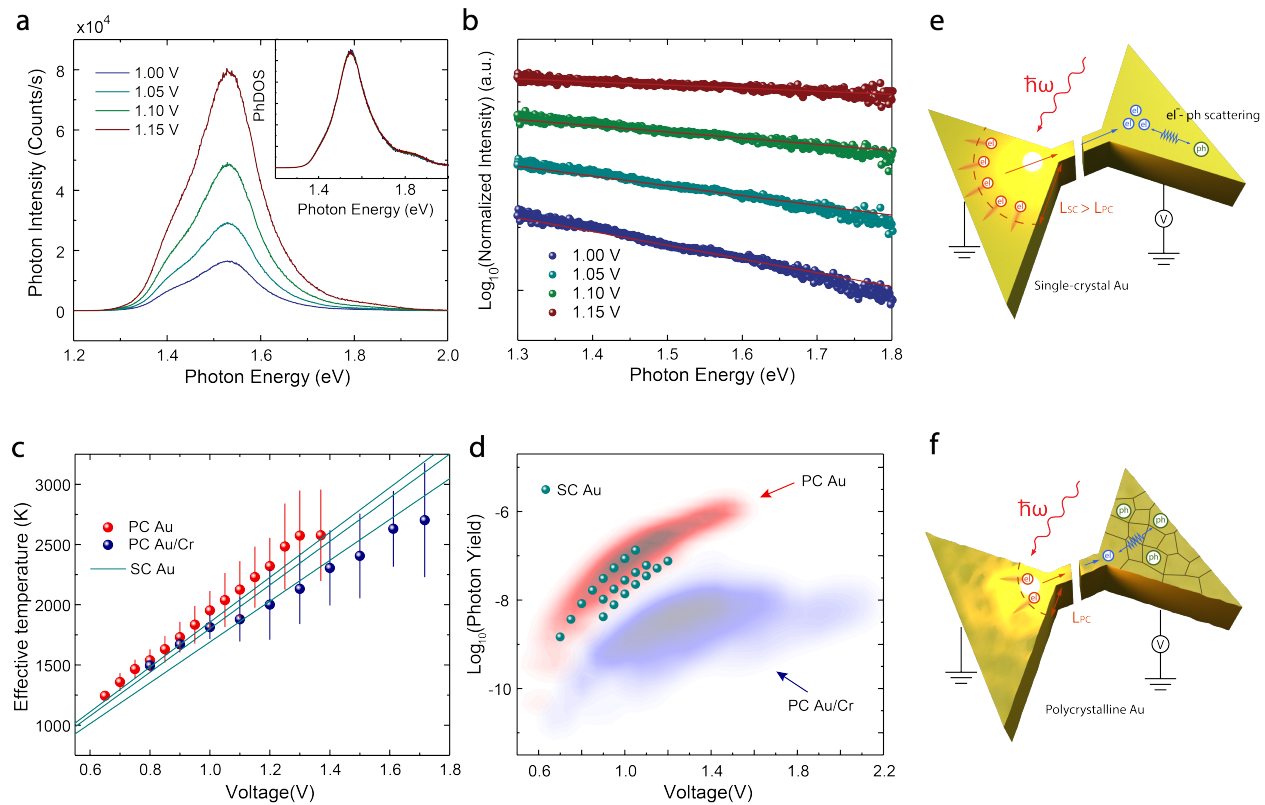

**Figure 4. Spectral measurements and analysis of electroluminescence from SC Au plasmonic nanogap devices and Schematics of physical mechanism underlying the observed large enhancement in hot carrier tunneling current.** (**a**) Measured light emission spectral at different bias. Inset shows the extracted surface plasmon modes through a normalization analysis. (**b**) Normalized spectra plotted on logarithmic scale versus the photon energy. Extracted effective temperatures of the hot carriers are shown in the legends. (**c**) Extracted effective temperatures vs. applied voltage bias for three different materials. (**d**) Photon yield comparison between SC Au (points) and PC Au and PC Au/Cr (density plots). (**e**) SC Au showing defect-free nanostructures that suppress electron-phonon scattering and increase the mean free path ($L_{SC}$) of hot carriers. (**f**) PC Au showing reduced hot carrier mean free path ($L_{PC}$) due to enhanced electron-phonon scattering.

The effective temperature extracted from the analysis in Fig. 4 provides valuable information regarding the energy distribution of the generated hot carriers in the plasmonic nanodevices. For a detailed comparison, the effective temperature as a function of applied bias, obtained from three different SC Au tunnel junctions, is plotted together with PC Au and Au/Cr tunnel junctions (Fig. 4c). The results show that SC Au nanogaps exhibit similar effective temperatures to PC Au, with both materials surpassing PC Au/Cr, suggesting that SC Au and PC Au have similar hot carrier energy distributions under identical excitation conditions. In addition, the photon emission yield (defined as the ratio of the photon counts to tunneling current) among different materials, which quantifies both the number density and energy of hot carriers, is shown in Fig. 4d. SC Au devices fall within the same photon yield range as PC Au device. The difference in photon yield between those and PC Au/Cr-based nanogaps can be attributed to Cr acting as a plasmonic damping medium to suppress the generation of hot carriers and lower their average energy.

It should be noted that the above analysis leads to the following interesting conclusion: SC and PC Au nanogap devices exhibit similar hot carrier generation efficiency and effective temperatures. This is in fact counterintuitive because one may expect that plasmonic field enhancement together with the differences in the electronic structure can affect the rates of plasmon-induced hot carrier excitation and thus the hot carrier dynamics. If the average energy of the hot carrier ensemble rises, the corresponding tunneling current will increase as well due to the increased energetic breadth of the electronic distribution and the energy-dependent tunneling transmittance, leading to a higher OCPV signal. However, in conjunction with the fact from Fig. 3 that the plasmonic field enhancement in both types of devices cannot solely account for the observed large enhancement in OCPV signal, the high performance photovoltage detection capability of SC Au nanogap devices is most likely due to the difference in hot carrier transport properties within the metal, where the increased crystallinity and uniformity of SC Au can result in a longer hot carrier mean free path and thus increase the number of carriers reaching the nanogap for tunneling (see Fig. 4e-f for the schematics and below for the detailed discussion).

**Discussion**

Noted that a rigorous, quantitative modeling of the above physical mechanism would involve solving complex time-dependent Boltzmann transport equations or performing nonequilibrium Green's function (NEGF) calculations to capture the effects of material- and geometry-dependent mean free paths on hot carrier distributions within the plasmonic tunnel gap[56–58]. Such calculations are beyond the scope of this work. However, here we show that it is possible to qualitatively estimate the relative electron-phonon relaxation time and hot carrier mean free path between the PC and SC Au within the nano constriction area based on the enhancement ratio of the hot carrier tunneling current, thus establishing the nanogap devices as simple measurement platforms to infer transport properties of nonequilibrium hot carriers at their intrinsic interaction scale. In future works that use rigorous theoretical treatment to capture the full spatio-temporal evolution of non-equilibrium carriers, NEGF methods can be utilized to simulate the quantum transport of hot carriers across the nanogap under realistic bias and illumination conditions. These methods are well suited for modeling energy- and momentum-resolved carrier dynamics in open quantum systems, and could provide detailed insight into how device geometry and crystallinity influence tunneling probability and carrier relaxation. Additionally, Monte Carlo simulations could be employed to statistically evaluate carrier trajectories and scattering events within single- and polycrystalline structures, especially to resolve the interplay between elastic/inelastic scattering and geometric boundary conditions. These advanced computational techniques, integrated with full-field electrodynamics simulations of plasmonic enhancement, would enable predictive modeling of hot-carrier behavior in complex plasmonic architectures, and help bridge the gap between experimental results and microscopic electron transport processes.

The hot carrier dynamics in SC Au and PC Au nanostructures can be treated as two coupled systems: the electron and phonon baths. Initially, optical excitation generates non-thermalized hot electrons, which rapidly undergo electron-electron (el-el) scattering within 10s to 100s of fs, forming a steady-state distribution that can be described by a high effective temperature that can be obtained using our normalization analysis described above. Following el-el scattering, electron-phonon (el-ph) scattering occurs, relaxing the hot electron energy to the lattice on the ps timescale. In a simple model that neglects spatial diffusion of carriers and their energy, the entire process can be described by three coupled differential equations[31], with the energy transfer from electrons to the lattice using a modified Tomchuk-Fedorovich approach[59], which includes the additional boundary scattering within the nano constriction. Substituting the characteristic length ($L = 700$ nm, determined by the laser spot size and the nanogap geometry) and the el-el scattering rate for SC Au and PC Au (18 THz and 24 THz that are obtained in previous measurements[31]), the relative effective temperature difference of hot carriers in SC and PC Au devices is estimated to be less than 6%, which is consistent with our normalization analysis of the EL spectrum (see Fig. 3 and Supporting Information).

As shown in Fig. 3e-f, the OCPV in these nanogaps is due to the developed potential within the gap, which balances the net tunneling current of hot carriers. This tunneling current $I_{hc}$ can be expressed using the Landauer theory[52]:

$$I_{hc} = \frac{2e}{h} \int (f_L(E) - f_R(E)) T(E) dE \qquad (1)$$

where $T(E)$ is the energy dependent transmission function, $f_{L/R}(E)$ represents the electron distribution on the left and right electrode of the nanogap, respectively. To quantify the enhanced carrier collection efficiency, we need to estimate the number of hot electrons that can reach the nanogap before scattered by the boundary or lattice defects. Due to boundary scattering within the narrow nanowire region, we estimated the collection efficiency of hot carriers by assuming a small angle sector, with its radius directly proportional to the hot carrier mean free path (inverse to the carrier lifetime.). Here, by referring to the previous pump probe measurements for the hot carrier lifetime in SC (3.1 ps) and PC (1.6 ps) [60] Au nanoparticles, one can obtain an enhancement ratio of ~4, which is close to our measured value of 4.5 in tunneling current. In addition, the resistivity value of the SC and PC Au nanowires have also been measured as a function of temperature and shown in Fig. S3, which qualitatively estimates the carrier mean free path near the Fermi level and is consistent with our OCPV enhancement. Combining these above estimates, our experimental results and dynamic modeling suggest that the enhanced performance of SC Au compared to PC Au tunnel junctions for hot electron tunneling and optoelectronic conversion have been primarily attributed to reduced electron phonon scattering, which allows hot electrons to retain their energy for longer periods (i.e., longer mean free path, $L_{SC} > L_{PC}$), thereby increasing the tunneling current significantly. Further, given the simplicity of our plasmonic nanodevices, our analysis has also established the feasibility of leveraging the combination of experimental characterizations (transport measurements, electroluminescence spectroscopy, OCPV scans) to probe nanoscale hot carrier transport properties. Particularly, spectroscopic measurements effectively reveal the hot carrier effective temperature and steady state energy distributions. Photon yield analysis can infer the relative density of the hot electrons and hot holes within the nanogap area. By comparing the hot electron tunneling current inferred from the OCPV measurements, we can estimate the geometrically mediated material-dependent hot carrier lifetime and mean free path. These quantities are very challenging to obtain directly so far through ultrafast pump-probe techniques.

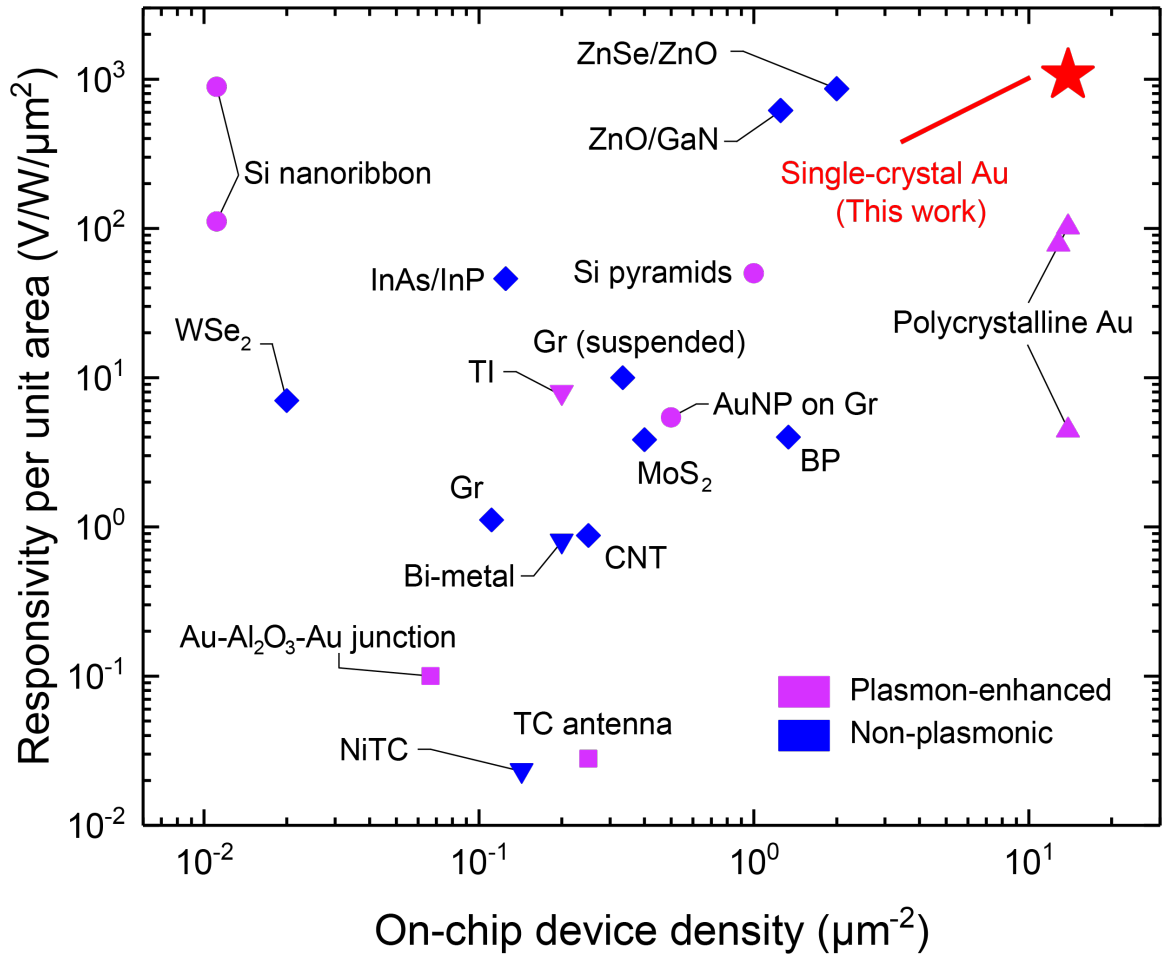

**Figure 4. Responsivity and on-chip device density of different plasmon-enhanced and non-plasmonic photovoltage detection methods.** The plasmon-enhanced photodetectors include Si nanoribbon at high(top) and low(bottom) optical power[61], Si pyramids [62], Au nanoparticles (AuNP) on graphene (Gr)[63], planar metal–oxide–metal (Au-Al$_2$O$_3$-Au) tunnel junction)[64], Thermocouple (TC) antenna)[65], Topological insulators (TI)[66], and polycrystalline Au plasmonic nanogaps [52]. The non-plasmonic photodetectors include transition metal dichalcogenide WSe$_2$[67] and MoS$_2$[68], black phosphorus (BP)[69], supported[70] and suspended graphene[71], InAs/InP nanowire[72], suspended carbon nanotubes (CNT)[73], bi-metal nanowire thermocouple (Bi-metal)[74], single metal (NiTC) nanostructure[75], ZnSe–ZnO nanowire axial p–n junctions[76] and ZnO/GaN nanoscale p-n junctions[77].

Following our fundamental understanding of the high performance photovoltage detection in SC Au nanogap devices, it is of great interest to compare the performance of the novel hot-carrier tunneling based mechanism with other reported schemes. We have shown in Fig. 4 the advantage of SC Au devices as ultraminiaturized on-chip photodetector by surveyed the responsivity per unit area and the integrated on-chip device density (showing the capability for large scale integration) in previous literature. It should be clarified that the area here is defined as the minimum functional unit necessary to replicate the photodetector structure for integration into larger array designs. In our case, this corresponds to the nanowire dimensions, as nanofabrication techniques can be utilized to fabricate array structures with numerous parallel nanowires sharing common source and drain pads. Two main types of photodetectors, plasmon-enhanced and non-plasmonic photodetectors, are shown for comparison. Here we focus on the photodetectors operating in open-circuit photovoltage mode without external load resistance, which is the same as the SC and PC Au devices. A number of short circuit photocurrent mode photodetectors that can be converted into photovoltage mode based on the photodiode current-voltage characteristic and internal resistance have also been summarized. As can be clearly seen, the SC Au devices exhibit excellent performance in both responsivity and device density, representing one of the most compact and sensitive photodetectors to the best of our knowledge.

While this work focuses on SC Au nanostructures, the underlying enhanced hot carrier transport mechanism (reduced electron-phonon scattering and longer mean free paths) is expected to be broadly applicable to other SC plasmonic metals, such as Ag and Al. Silver has even lower intrinsic losses and higher conductivity than gold, which could further improve both plasmonic field enhancement and hot carrier mean free path, potentially yielding even stronger photovoltage responses. Aluminum, although plasmonically active in the UV-visible range and attractive due to its abundance, presents challenges due to its rapid surface oxidation, which can severely degrade plasmonic performance and carrier transport at the metal–gap interface. However, such limitations may be addressed by incorporating ultrathin dielectric passivation layers (e.g., $Al_2O_3$ or $SiO_2$) via uniform deposition processes such as atomic layer deposition. These strategies are commonly used in MEMS processes and could stabilize surface properties without significantly modifying electron tunneling or optical coupling. Therefore, the generalization of SC plasmonic optoelectronic devices to other materials holds promise for expanding spectral tunability, improving cost-efficiency, and enabling compatibility with diverse photonic platforms. These directions can be possible research directions in future studies.

In this work we studied both the fundamental and applied plasmon-enhanced optoelectronics of single crystal metal nanostructures. A series of optoelectronic functions were experimentally studied including nanoscale hot carrier transport, electroluminescence, and photovoltage detection. Notably, our experimental results show dramatic enhancement in the generated open-circuit photovoltage in SC plasmonic nanodevices, which is primarily attributed to the enhanced hot carrier transport properties, rather than the plasmonic field enhanced effects. The combination of reduced hot carrier scattering and optimized nanogap geometry in SC Au nanostructures offers a promising route toward improving the performance of on-chip optoelectronic and nanophotonic devices. We anticipate that the single-crystal nanoscale platform in this work can be expanded to the studies of various electrically driven nanophotonic phenomena and applications such as electrically-driven plasmon-mediated photocatalysis, strong coupling, and atomic-sized light sources.

**Materials and Methods**

**Chemical synthesize of SC Au flakes**

The procedure to synthesize the gold-flakes is adapted from previous study[78]. For our experiments, a 20 mL ethylene glycol growth solution and 180 μL of HPLC water are added to 7 mg of Gold(III) Chloride salt. The solution is heated to 80°C in a water bath, and 90 μL aniline is gradually added during the heating process. The solution is then quickly moved into a dark vacuum rotary oven at 90°C. The gold flakes can be observed after 9-18 hours of incubation. The solution is replaced with absolute ethanol using a vacuum filtration system to clean the flakes. Once ethanol is flushed, the gold aggregates and unwanted smaller particles are removed using a 2 μm poly-propylene syringe filter. The desired flakes are obtained by reverse flushing UHP nitrogen through the filter.

**Formation of nanogap devices using electromigration break junction technique**

The electromigration break junction process begins by applying slow voltage sweeps (10 mV/s) to the nanowires using a Keithley 2400 source meter. The electrical current is continuously monitored so that the voltage sweep can stop immediately when a sudden drop (~0.4-2.0%) in current is detected. This drop indicates a slight resistance increase due to atomic migration under the electrical field within the nanowire. Additional voltage sweeps are then applied to further induce electromigration until the resistance reaches desired value (0.05 $G_0$ to 0.1 $G_0$)[47]. This method is essential for achieving a high yield in forming stable tunnel junctions, which are critical for OCPV and light emission measurements.

**Optical measurement setup**

A linearly polarized 785 nm CW laser (aligning to the resonant wavelength of the transverse plasmon mode within the Au nanowire) is focused through a high NA objective (Nikon 50×, NA 0.7) with a diameter of 1.8 μm onto the junction. The laser is then raster scanned across the nanowire through a nanopositioner (ANC 300 Piezo Controller) while being chopped at a frequency synced to the lock-in amplifier. The open circuit photovoltage is then amplified through a voltage preamplifier (SR560) and collected by the lock-in amplifier. The electroluminescence spectra from the biased tunnel junction are measured using a home-built Raman setup, consisting of free space optical elements and a CCD spectrometer (Horiba iHR 320/Synapse CCD).

## Acknowledgments

L.C. and Y.C. acknowledge support from Air Force Research Laboratory (AFRL FA8651-22-1-0012) and National Science Foundation (NSF CAREER CBET-2239004). D.N. and S.L. acknowledge support from the Office of Naval Research (N00014-21-1-2062), the National Science Foundation (ECCS-2309941), and the Robert A. Welch Foundation (C-1636).

**Author contributions:**

L.C.,Y.Z. and D.N. designed the experiment. Y.Z. fabricated the devices, conducted the experiments. S.C.Y. and J.S. synthesized the single crystal gold flakes. S.L. performed the finite-element simulation. L.C. and Y.Z wrote the manuscript with comments and inputs from all authors.

**Data and materials availability:**

The data that support the findings of this study are available from the corresponding author upon reasonable request.

**Competing interests:**

The authors declare that they have no competing interests.

**Supplementary information**

Supplementary information has been provided in the separated file.

# Supplementary Materials for

## On-chip Single Crystal Plasmonic Optoelectronics for Efficient Hot Carrier Generation and Photovoltage Detection

### 1. A simplified model of carrier temperatures

A full theoretical model of the nanogap under illumination and/or bias is extremely challenging, requiring a treatment of the local plasmonic excitations, the spatially and temporally dependent electronic distribution, and the open nature of the system, including diffusion of the carriers and the energy content. Inferring an open-circuit photovoltage would further require some definition of the asymmetry of the junction (either in the generation of hot carriers or the tunneling kinetics) that leads to a net hot-carrier tunneling current under illumination. Such a model is well beyond the scope of the present work. However, some insight can be gained by considering a much-simplified treatment, adapted from approaches applied to nanoparticles, that treats the energy density, effective temperature of a population of locally thermalized electrons, and the lattice temperature as *local* variables.

After the electrical or optical excitation of the hot carriers, within the first tens to hundreds of femtoseconds, excited nonthermalized hot electrons undergo rapid electron-electron scattering, quickly relaxing their energy and establishing a steady-state hot carrier distribution characterized by a high effective temperature. Electron-phonon scattering subsequently comes into play to further relax the hot carrier energy to the lattice on the picoseconds time scale. The whole process in this simplified system can be well described by three coupled differential equations(*1*),

$$\frac{\partial N(t)}{\partial t} = -\gamma_{e-e}N(t) - \frac{G_{e-ph}}{C_e}N(t) + P_{ex}(t) \quad (1)$$

$$C_e \frac{\partial T_{eff}(t)}{\partial t} = -G_{e-ph}\left(T_{eff}(t) - T_l(t)\right) + \gamma_{e-e}N(t) \quad (2)$$

$$C_l \frac{\partial T_l(t)}{\partial t} = G_{e-ph}\left(T_{eff}(t) - T_l(t)\right) + \frac{G_{e-ph}}{C_e}N(t) - P_{sub}(t) \quad (3)$$

where $N(t)$ is the energy density of the excited non-thermal electrons, $\gamma_{e-e}$ is the electron-electron scattering rate, $G_{e-ph}$ is the electron phonon coupling constant. $C_e, T_{eff}(t)$ and $C_l, T_l(t)$ are the heat capacity and temperature (effective temperature for steady state hot carriers) of electron and phonon respectively. $P_{ex}(t)$ and $P_{sub}(t)$ are the absorbed optical excitation power and energy relaxation rate to the external environment (substrate).

Augmenting this local model to account at least a bit for sample geometry, different from the previous study on large monocrystalline gold micro flakes, the bow-tie shaped constriction employed here will introduce more boundary scattering for the electrons and thus providing an additional relaxation channel that needs to be considered. Following the paper by Tomchuk and Fedorovich(*2, 3*), the energy transferred to the lattice can be expressed by,

$$P_{e-ph} = \Gamma(T_{eff}^2 - T_l^2) \tag{4}$$

Where $\Gamma$ is a factor that contains the characteristic length $L$ of the constriction area,

$$\Gamma = \frac{\pi^2}{4}\left(\frac{m^2 k_B^2}{\hbar^3 M}\right) E_F \frac{1}{L} \tag{5}$$

here $m$ and $M$ are the mass for electron and atom, $E_F$ is the Fermi energy, $k_B$ is the Boltzmann constant, and $\hbar$ is the reduced Planck's constant. Eq. (4) is then used to substitute for the electron phonon energy relaxation term in the bulk $G_{e-ph}\left(T_{eff}(t) - T_l(t)\right)$ in Eq. (5).

Under CW optical excitation, the system will eventually reach a steady state (the time-dependent terms in Eq. (1-3) all vanish), where the relative electron effective temperature difference can be compared by inserting the physical parameters into Eq. (4). According to previous studies(*1, 4*), $G_{e-ph}$ are taken as $2.2 \times 10^{16}$ Wm$^{-3}$K and $2.0 \times 10^{16}$ Wm$^{-3}$K for SC and PC gold respectively. $C_e$ is 67 Jm$^{-3}$K$^{-2}$ and $E_F$ is 5.52 eV for gold. Assuming one side is unsymmetrically illuminated for the optimal OCPV signal and a beam spot size of 2um, $L$ is taken to be 700 nm. Electron-electron scattering rate is taken as 18 THz and 24 THz for SC and PC from previous ultrafast pump-probe experiments(*1*). The relative effective temperature difference for SC and PC gold can then be estimated to be less than 6%, which is consistent with the extracted effective temperature value from the EL data. Thus, this simplified model is consistent with the experimental observation that effective temperatures in the steady state of SC and PC junctions are quite similar. As stated in the main text, this implies that the tunneling kinetics of hot carriers in the SC and PC junctions, explicitly neglected here, must be the origin of the large difference in OCPV between the SC and PC junctions.

## 2. Finite element simulation

A finite element method (FEM) simulation is performed to model the electromagnetic response of a gold nanowire system using COMSOL Multiphysics with the electromagnetic wave, frequency domain (ewfd) module. The whole system for modelling includes the gold nanowire (single crystal gold and evaporated gold), 2 μm thick SiO$_2$ substrate on Si and the vacuum above the nanowire. The values of dielectric function of single crystal gold and evaporated gold are obtained from the previous studies[55]. The material properties of the $SiO_2$ and Si are selected in the COMSOL material library. The whole geometry is surrounded by a perfect match layer (PML) to mimic an open and nonreflecting infinite domain. The laser is set to be a perfect Gaussian beam with the spot size to be 2 μm in radius used in the experiment[52]. The polarization angle of the laser is set to be either parallel (0 degree) or transverse (90 degree) to the nanowire. A minimum mesh size of 0.5 nm was selected to ensure numerical accuracy for resolving small gaps and interfaces, with at least two mesh elements being covered inside the gap. We first calculate the background field in the free space, and then we use the results to calculate the full scattered field to obtain the enhancement shown in the main text.

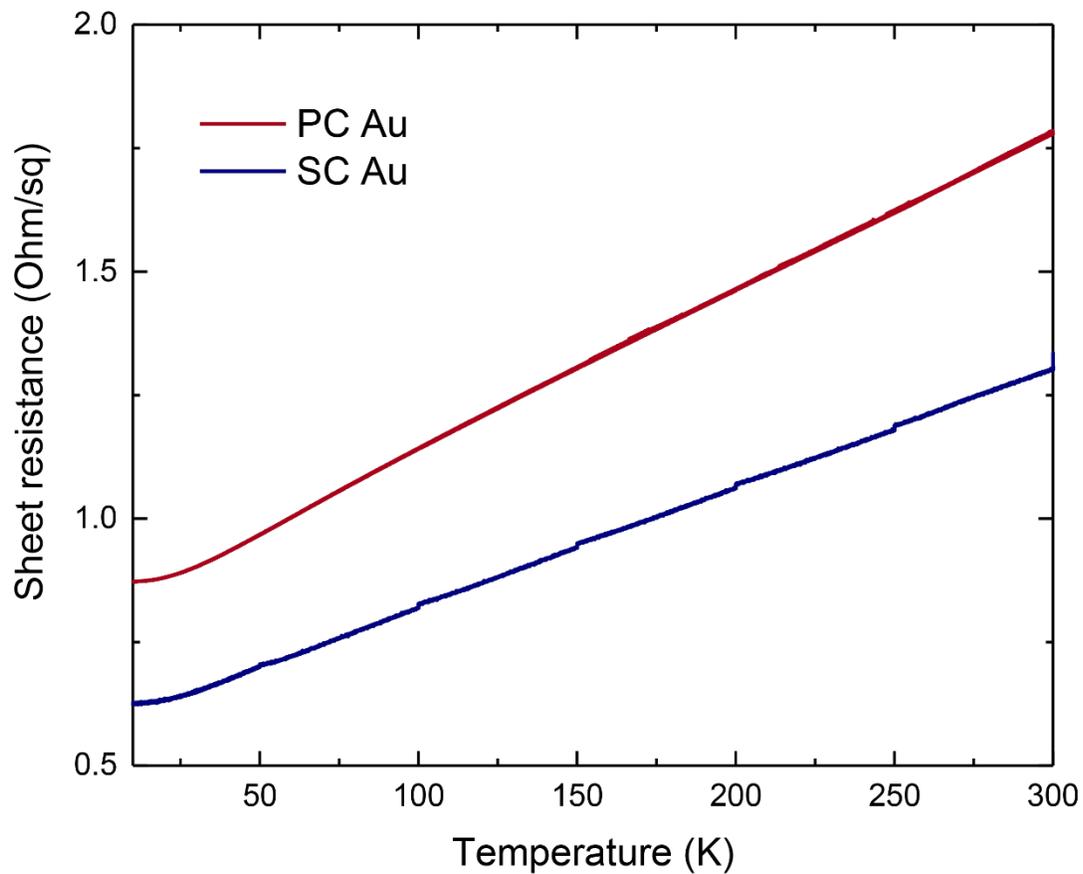

**Fig. S1.** Sheet resistance of 20nm thick gold as a function of temperature for PC and SC Au respectively. The geometry for PC Au nanowire is 100 μm long, 2um wide. The geometry for SC Au nanowire is 1.8 μm long, 280 nm wide.

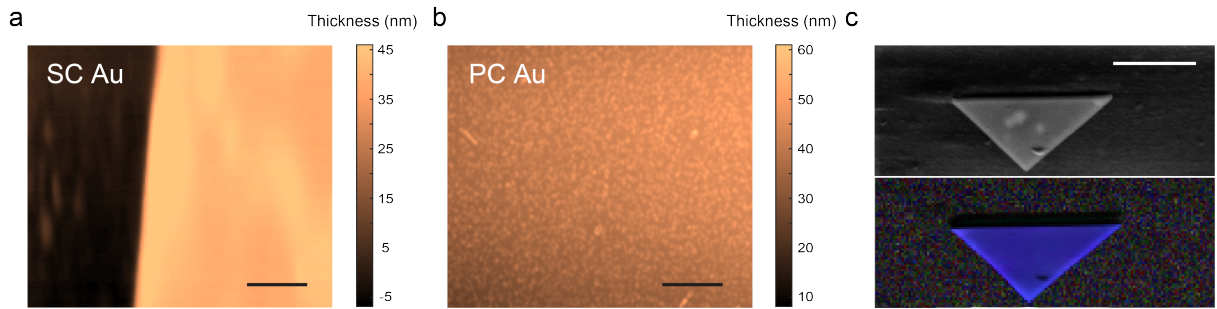

**Fig. S2.** (a-b) AFM scanning image for SC and PC Au respectively. The range of the color bar have been chosen to be the same for fair comparison. The scale bar in the figure is 2 **μ**m. (c) SEM image (top) and EBSD inverse pole figure (IPF) map (bottom) showing the crystallographic orientation of a single crystalline Au flake deposited on top of an e-beam evaporated polycrystalline Au film. The triangular region exhibits a uniform color, indicating a consistent crystallographic orientation corresponding to the Au (111) plane. In contrast, the surrounding evaporated Au shows multiple grain orientations and visible grain boundaries. The scale bar in the figure is 2 **μ**m.